\documentclass{pasj01}
\Received{October 5, 2021}
\Accepted{November 29, 2021}
\Published{$\langle$publication date$\rangle$}
\usepackage{scalefnt}
\usepackage{natbib}
\usepackage{url}
\usepackage{lineno}
\usepackage{longtable}
\usepackage{ulem} 

\begin{document}

\title{VERA astrometry toward the Perseus arm gap}

\author{
Nobuyuki \textsc{Sakai}\altaffilmark{1},
Hiroyuki \textsc{Nakanishi}\altaffilmark{2},
Kohei \textsc{Kurahara}\altaffilmark{3},
Daisuke \textsc{Sakai}\altaffilmark{4},
Kazuya \textsc{Hachisuka}\altaffilmark{3},
Jeong-Sook \textsc{Kim}\altaffilmark{1, 5},
and
Osamu \textsc{Kameya}\altaffilmark{3, 6}
}%
\altaffiltext{1}{Korea Astronomy $\&$ Space Science Institute, 776, Daedeokdae-ro, Yuseong-gu, Daejeon 34055, Korea}
 \altaffiltext{2}{Graduate Schools of Science and Engineering, Kagoshima
  University, 1-21-35 Korimoto Kagoshima}
 \altaffiltext{3}{Mizusawa VLBI Observatory, National Astronomical
   Observatory of Japan, 2-12 Hoshigaoka, Mizusawa-ku, Oshu, Iwate 023-0861, Japan}
\altaffiltext{4}{National Astronomical Research Institute of Thailand (Public Organization), 260 Moo 4, T. Donkaew, Amphur Maerim, Chiang Mai 50180, Thailand}
\altaffiltext{5}{Basic Science Research Institute, Chungbuk National University, Chungdae-ro 1, Seowon-Gu, Cheongju, Chungbuk 28644, Korea}
\altaffiltext{6}{Department of Astronomical Science, The Graduate University for Advanced Studies, 2-21-1 Osawa, Mitaka, Tokyo 181-8588, Japan}
 \email{nsakai@kasi.re.kr}


\maketitle

\begin{abstract}
The Perseus arm has a gap in Galactic longitudes ($l$) between 50$^{\circ}$ and 80$^{\circ}$ (hereafter the Perseus arm gap) where the arm has little star formation activity. To better understand the gap, we conducted astrometric observations with VERA and analyzed archival H {\scriptsize I} data. We report on parallax and proper motion results from four star-forming regions, of which G050.28-00.39 and G070.33+01.59 are likely associated with the gap. The measured parallaxes are 0.140$\pm$0.018 (mas), 0.726$\pm$0.038 (mas), 0.074$\pm$0.037 (mas), and 0.118$\pm$0.035 (mas) for G050.28-00.39, G053.14+00.07, G070.33+01.59, and G079.08+01.33, respectively. Since the fractional parallax error of G070.33+01.59 is large (0.5), we estimated a 3D kinematic distance of the source to be 7.7$\pm$1.0 kpc using both the LSR velocity ($V_{\rm{LSR}}$) and the measured proper motion. Perseus-arm sources G049.41+00.32 and G050.28-00.39 lag relative to a Galactic rotation by 77$\pm$17 km s$^{-1}$ and 31$\pm$10 km s$^{-1}$, respectively. The noncircular motion of G049.41+00.32 cannot be explained by the gravitational potential of the Perseus arm. We discovered rectangular holes with integrated brightness temperatures of $<$ 30 K arcdeg in $l$ vs. $V_{\rm{LSR}}$ of the H {\scriptsize I} data. One of the holes is centered near ($l$, $V_{\rm{LSR}}$) = (47$^{\circ}$, $-$15 km s$^{-1}$), and G049.41+00.32 is associated with the rim of the hole. However, G050.28-00.39 is not associated with the hole. We found extended H {\scriptsize I} emission on one side of the Galactic plane when integrating the H {\scriptsize I} data over the velocity range covering the hole (i.e., $V_{\rm{LSR}}$ = [$-$25, $-$5] km s$^{-1}$). G049.41+00.32 and G050.28-00.39 are moving toward the emission. The Galactic H {\scriptsize I} disk at the same velocity range showed an arc structure, indicating that the disk was pushed from the lower side of the disk. All the observational results might be explained by a cloud collision with the Galactic disk.


\end{abstract}


\section{Introduction}

The Perseus arm is one of the two (e.g., \citealp{2000A&A...358L..13D}; \citealp{2009PASP..121..213C}) or four (e.g., \citealp{1976A&A....49...57G,2003A&A...397..133R}) dominant spiral arms of the Milky Way, based on radio, infrared, and optical observations. The Red MSX Source survey \citep{2014MNRAS.437.1791U} showed that the arm had a large number of massive young stellar objects (MYSOs) at Galactic longitudes ($l$) between approximately 80$^{\circ}$ and 150$^{\circ}$, whereas a lower density of MYSOs was found at Galactic longitudes between approximately 50$^{\circ}$ and 80$^{\circ}$ (Fig. 14 of \citealp{2013ApJ...775...79Z}). A similar trend was found in the 1.1-mm dust continuum data tracing dense molecular cores (Fig. 6 of \citealp{2013ApJS..209....2S}) and the distribution of the Galactic H {\scriptsize II} regions \citep{2017PASP..129i4102K}. The star formation activity of the Perseus arm at the gap region (i.e., $l$ = [50$^{\circ}$, 80$^{\circ}$]; hereafter the Perseus arm gap) is lower, and this situation will continue in the future (i.e., lack of raw material of present and future star formation in the Perseus arm gap). 

In addition, 41 VLBI astrometric results tracing high-mass star-forming regions (HMSFRs), have been reported for the Perseus arm, of which only two are associated the Perseus arm gap (see Fig. 2 of \citealp{2019AJ....157..200Z}). For the Perseus arm at Galactic longitudes of less than 50$^{\circ}$, there are 11 VLBI astrometric results, among which the most luminous (10$^{7.2}$ solar luminosity ($L_{\odot}$) at a distance of 11.11 kpc; \citealp{1991A&A...251..231S,2013ApJ...775...79Z}) Galactic star-forming region W49N is included. Astrometric results for the Perseus arm gap should be increased for a better understanding of the arm.

The Perseus arm is also known to be surrounded by intermediate-velocity clouds (IVCs with $|V_{\rm{LSR}}|$ = 50$-$100 km s$^{-1}$) and high-velocity clouds (HVCs with $|V_{\rm{LSR}}|$ $>$ 100 km s$^{-1}$), although distance determinations for IVCs/HVCs have been limited (e.g., \citealp{1975A&A....40....1H}). \citet{2008ApJ...672..298W} conducted (optical) absorption line observations toward Galactic halo stars to constrain the distances of three HVCs and one IVC. They revealed that the IVC centered near ($l$, $b$) = (145$^{\circ}$, $-$42$^{\circ}$) is located 0.7$-$1.8 kpc below the Perseus arm. The IVC was regarded as the return flow of the Galactic fountain. Additionally, the Smith cloud (SC; \citealp{1963BAN....17..203S,2008ApJ...679L..21L}), a gaseous HVC, is located 2.9 kpc below the Perseus arm at ($l$, $b$, $V_{\rm{LSR}}$) = (38$^{\circ}_{.}$67, $-$13$^{\circ}_{.}$41, 99 km s$^{-1}$). The coordinates represent the tip of the cloud. The cloud is thought to have passed through the outer Galactic disk approximately 70 Myr ago, and is approaching the disk with a vertical velocity of 73$\pm$26 km s$^{-1}$ \citep{2008ApJ...679L..21L}. Although the metal-enriched nature of the SC favors a Galactic origin \citep{2016ApJ...816L..11F}, its large mass ($>$ 10$^{6}$ $M_{\odot}$) and prograde kinematics with $V_{\phi}$ = 270$\pm$21 km s$^{-1}$ at the tip remain to be explained (where $V_{\phi}$ is the azimuthal velocity in Galactocentric cylindrical coordinates). 

As mentioned above, the SC is thought to have survived the first disk transit although a cloud is destroyed during a transit \citep{2009IAUS..254..241B}. To explain the discrepancy, \citet{2016ApJ...816L..18G} and \citet{2018MNRAS.473.5514T} proposed that the SC is confined by a DM subhalo. Their simulation results showed that material removal due to ram pressure stripping is weakened thanks to the DM subhalo. A cloud's collision with the disk punched a hole in the disk in the simulation of \citet{2016ApJ...816L..18G}, whereas a cloud did not ``punch'' an apparent hole in the disk nor did it create a bubble in the simulation of \citet{2018MNRAS.473.5514T}. The difference might be due to assumed physical parameters in the simulations (e.g., the gas column density of the cloud relative to that of the disk at the impact point).

In this paper, we report on 22-GHz astrometric results for four sources expected to be associated with the Perseus arm gap. In Section 2, we explain VLBI data obtained by VERA observations. We present VLBI astrometric results in Section 3. The structure and kinematics of the Perseus arm gap are discussed in Section 4. In Section 5, we summarize the paper. 

Throughout the paper, we assume an A5 model of \citet{2019ApJ...885..131R}: Galactic constants of ($R_{0}$ = 8.15 kpc, $\Theta_{0}$ = 236 km s$^{-1}$); the ``universal'' form for the rotation curve (\citealp{1996MNRAS.281...27P}); solar motion of ($U_{\odot}$, $V_{\odot}$, $W_{\odot}$) = (10.6, 10.7, 7.6) km s$^{-1}$.

\begin{table*}[htbp] 
\caption{Observational Information\footnotemark[$*$]. \hspace{10em}} 
\begin{center} 
\label{table1} 
\small 
\begin{tabular}{lllccc} 
\hline 
\hline 
Source&R.A. &Decl. 	  &$\Delta\nu$&$\Delta B_{\rm{IF}}$/$N_{\rm{IF}}$ &Obs. Date in UT 	  \\
	       &hh:mm:ss &dd:mm:ss &(kHz)&(MHz)&20yy/mm/dd\\ 
\hline
G050.28-00.39	&19:25:17.9310  	&+15:12:24.720	&31.25			&16/1&A. 16/11/25; B. 17/01/31; C. 17/04/02; D. 17/05/29	\\                                              
 J1924+1540	&19:24:39.4559    	&+15:40:43.942&500	&16/15\footnotemark[$\dag$]&E. 17/09/12; F. 17/11/20; G. 18/01/27; H. 18/03/02  \\   
\multicolumn{3}{l}{(SA = 0.5$^{\circ}$; PA = -18$^{\circ}$)\footnotemark[$\ddag$]} & &&I. 18/04/07; J. 18/05/19 \\ 
 J1930+1532	&19:30:52.7670    	&+15:32:34.427 	&500&16/15\footnotemark[$\dag$]	\\
\multicolumn{3}{l}{(SA = 1.4$^{\circ}$; PA = 76$^{\circ}$)\footnotemark[$\ddag$]} \\ \\
G053.14+00.07	&19:29:17.4764  	&+17:56:18.264&15.625&16/1&A. 18/11/07; B. 19/01/12; C. 19/03/16; D. 19/05/27 \\
J1927+1847&19:27:32.3124&+18:47:07.902 &250 &16/15\footnotemark[$\dag$]&E. 19/08/31; F. 19/11/16; G. 20/01/16; H. 20/03/05 \\
\multicolumn{3}{l}{(SA = 0.9$^{\circ}$; PA = -26$^{\circ}$)\footnotemark[$\ddag$]} &&&I. 20/05/18; J. 20/06/18 \\
J1928+1842	&19:28:54.9610    	&+18:42:31.241&250&16/15\footnotemark[$\dag$]\\
\multicolumn{3}{l}{(SA = 0.8$^{\circ}$; PA = -7$^{\circ}$)\footnotemark[$\ddag$]}&&	\\ \\
G070.33+01.59	&20:01:54.0627  	&+33:34:14.485&15.625&16/1&A. 18/11/01; B. 19/01/06; C. 19/03/18; D. 19/05/30	\\ 
J1957+3338	&19:57:40.5499    	&+33:38:27.943&1,000 &512/1&E. 19/09/02; F. 19/11/25; G. 20/01/18; H. 20/02/21 \\
\multicolumn{3}{l}{(SA = 0.9$^{\circ}$; PA = -85$^{\circ}$)\footnotemark[$\ddag$]}							 &&			&I. 20/04/29; J. 20/06/03 \\
J2010+3322	&20:10:49.7231    	&+33:22:13.812	&1,000	&512/1 \\
\multicolumn{3}{l}{(SA = 1.9$^{\circ}$; PA = 96$^{\circ}$)\footnotemark[$\ddag$]} \\ \\
G079.08+01.33	&20:27:20.2734  	&+40:42:34.648 &31.25&16&A. 15/12/25; B. 16/02/25; C. 16/04/22; D. 16/10/09\\
J2033+4000	&20:33:03.6706    	&+40:00:24.414 &500&16/15\footnotemark[$\dag$]&E. 16/11/27; F. 17/01/07; G. 17/03/10; H. 17/05/06 \\
\multicolumn{3}{l}{(SA = 1.3$^{\circ}$; PA = 122$^{\circ}$)\footnotemark[$\ddag$]} &&&I. 17/09/09 \\

\hline 
\multicolumn{4}{@{}l@{}}{\hbox to 0pt{\parbox{165mm}{\normalsize
\par\noindent
\\
Column 1: 22 GHz H$_{2}$O maser source (as denoted by ``G") and background
QSOs (as denoted by ``J"); Columns 2-3: equatorial coordinates in (J2000); Column 4: frequency spacing; Column 5: bandwidth and number of IF; Column 6: observation date in UT. \\
\footnotemark[$*$] Coordinate values of maser sources described here, were derived by the phase referencing relative to individual background (continuum) sources. These values are used as individual origins of maser distribution maps in Appendix Fig. \ref{fig:5}.\\
\footnotemark[$\dag$] Fifteen IFs span 464 MHz with a space of 16 MHz among individual IFs.\\
\footnotemark[$\ddag$] Separation Angle (SA) and Position Angle (PA) East of North of the background QSO with respect to the target (maser) source.
}\hss}}
\end{tabular} 
\end{center} 
\end{table*}

\section{Observations}
Candidates of the Perseus-arm source hosting 22-GHz water masers were selected based on maser (\citealp{2001A&A...368..845V}; \citealp{2007PASJ...59.1185S}; \citealp{2016ApJ...822...59S}) and molecular lines catalogs (\citealp{2002ApJS..141..157Y}; \citealp{2013ApJS..209....2S}) as well as the position velocity diagram of CO (that is $l-v$ plot; \citealp{2001ApJ...547..792D}). VLBI astrometric observations at 22 GHz were performed for the candidates with VERA. Techniques for VERA astrometry are summarized in previous VERA astrometry papers (e.g., \citealp{2012PASJ...64..108S}). The observational information is summarized in Table \ref{table1} where coordinate values of the maser sources are not identical to those used for the observations. This is because we could accurately determine the masers' positions by the phase referencing relative to background (continuum) sources. We used the VERA dual-beam system (\citealp{2000SPIE.4015..544K}) to simultaneously observe the target and the phase reference, which enabled us to compensate for the atmospheric phase fluctuation for high-precision astrometry. To calibrate the clock parameters, a bright continuum source was observed for 5 min every 80 min in the observations. 

After filtering was performed with the VERA digital filter (\citealp{2005PASJ...57..259I}), left-handed circular polarization data, except for G070.33+01.59 data, were recorded at 1,024 Mbps with 2-bit quantization. The data of G070.33+1.59 and adjacent phase reference were recorded at 2 Gbps. The details of the VERA back-end system are shown in Fig. 2 of \citet{2016PASJ...68..105O}. All data sets were correlated with the Mizusawa software correlator\footnote{\url{https://www.miz.nao.ac.jp/veraserver/system/fxcorr-e.html}}. The frequency spacing and bandwidth of the maser and continuum data are listed in Table \ref{table1}.

\section{Data reduction}
Data reduction was performed with the Astronomical Image Processing System developed by NRAO ($AIPS$; \citealt{1996ASPC..101...37V}). We employed the general procedure utilized in previous VERA astrometry papers (e.g., Fig. 11 of \citealp{2011PASJ...63..513K}; Fig. 5 of \citealp{2012PASJ...64..142I}; \citealp{2020PASJ...72...51N,2020PASJ...72...52N}). By considering the flux density and compactness of the maser and continuum sources (i.e., background QSOs), we chose a bright and compact source as a phase reference, and the adjacent source (maser or QSO) was imaged through phase referencing.  

The relative position obtained through the phase referencing was recorded as a function of time during the observations. The resultant position offsets of the maser source were modeled using (1) the annual parallax, (2) linear proper motion components in the east (R.A.cos$\delta$) and north ($\delta$) directions, and (3) reference positions of R.A.cos$\delta$ ($\alpha_{0}$) and $\delta$ ($\delta_{0}$) (e.g., see Eq. 1 of \citealp{2009ApJ...694..192M}). The model parameters were determined according to the weighted least squares, in which a systematic error was added in quadrature to the formal (thermal) error so that the reduced chi-square value nearly reached unity. This is because systematic errors caused by tropospheric zenith delay residuals are dominant in 22-GHz VLBI astrometry (e.g., see \citealp{2014ARA&A..52..339R}).

\begin{figure*}[tbhp] 
 \begin{center} 
     \includegraphics[scale=1.1]{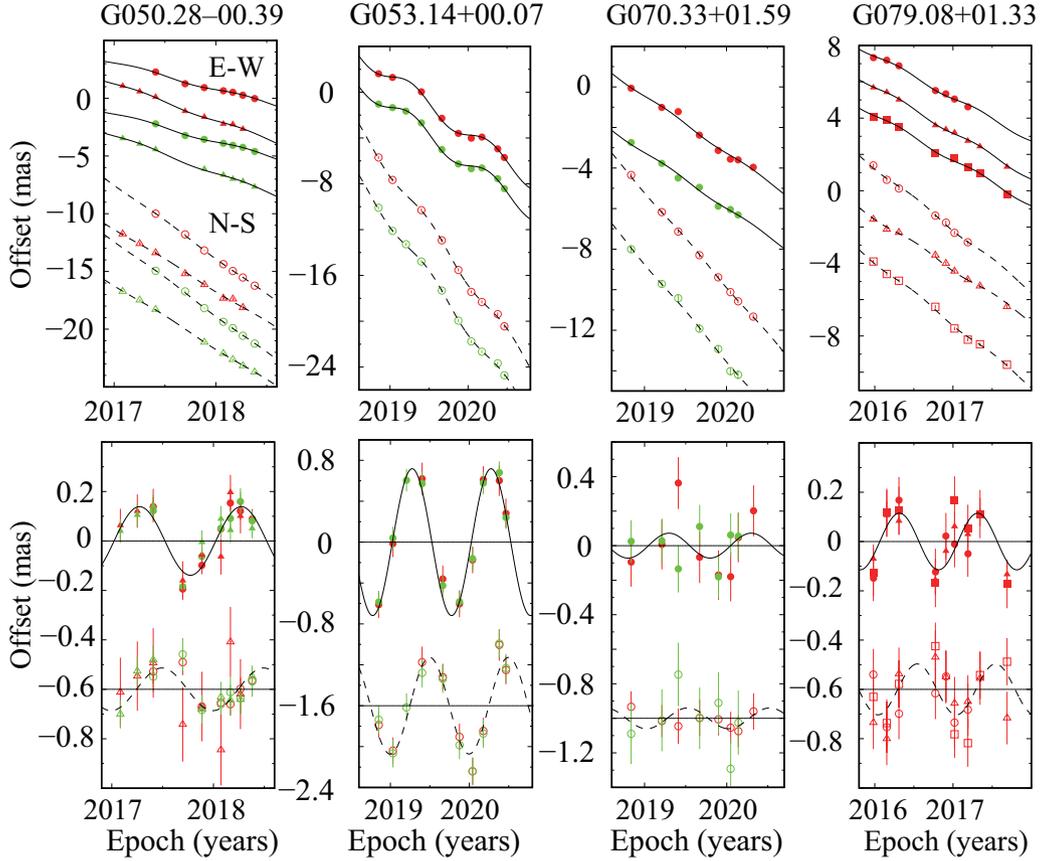} 
\end{center} 
\caption{ Results of parallax and proper-motion 
fitting. Plotted are position offsets of maser spots with respect to background
QSOs toward the east and north as a function of time. For clarity, the data toward the north direction is plotted offset from the
 data toward the east direction. Different color and symbol separate individual background QSOs and maser features, respectively (see Appendix Table \ref{table:3}). \textbf{(Top row)} The best-fit models in the east and north directions are shown
as continuous and dashed curves, respectively.  
\textbf{(Bottom row)} Same as top row, but with proper motions removed (i.e., parallax results).}
\label{fig:1} 
\end{figure*}  

\begin{table*}[htbp] 
\caption{Parallax and proper motion results. \hspace{10em}} 
\begin{center} 
\label{table:2} 
\small 
\begin{tabular}{llcccrc} 
\hline 
\hline 
 Target			&Parallax ($\pi$)&Distance &$\mu_{\alpha} \rm{cos}\delta$	&$\mu_{\delta}$&$V_{\rm{LSR}}$		&Ref.	 \\

	&(mas)&(kpc)&(mas yr$^{-1}$)	&(mas yr$^{-1}$)&(km s$^{-1}$)						\\
  \hline
G050.28-00.39					&0.140$\pm$0.018			&7.1$^{+1.1}_{-0.8}$		&$-$3.29$\pm$0.56	 &$-$5.52$\pm$0.37	 	&17$\pm$3&a	    		\\
G053.14+00.07					&0.726$\pm$0.038			&1.4$\pm$0.1		&$-$1.27$\pm$1.08	 &$-$7.15$\pm$1.07	 		    	&22$\pm$4&a	\\

G070.33+01.59					&0.074$\pm$0.037\footnotemark[*]			& \ \ \ $-$	&$-$2.82$\pm$0.29	 &$-$4.68$\pm$0.28	 		    	&$-$23$\pm$5&b	\\

G079.08+01.33					&0.118$\pm$0.035			&8.5$^{+3.6}_{-1.9}$	&$-$2.49$\pm$0.14	 &$-$3.36$\pm$0.24	 &$-$18$\pm$5\footnotemark[\dag], $-$64$\pm$1\footnotemark[\dag]&c, d		    		\\

\hline 
\multicolumn{4}{@{}l@{}}{\hbox to 0pt{\parbox{140mm}{\normalsize
\par\noindent
\\
Column 1: 22 GHz H$_{2}$O maser source; Columns 2-3: parallax and corresponding distance; Columns 4-5; proper motion components in east and north directions, respectively; Column 6: LSR velocity; Column 7: reference for the LSR velocity estimated from a molecular line observation. \\
$\bf{References}$: (a) \citet{2013ApJS..209....2S}; (b) \citet{1996ApJ...463..205A}; (c) \citet{2002ApJS..141..157Y}; (d) \citet{2017ApJ...834...57M}. \\
\footnotemark[*] Since the fractional parallax error is 50$\%$, distance estimation by simply inverting the parallax results in a significant bias (see \citealp{2015PASP..127..994B}).   \\
\footnotemark[\dag] Different LSR velocities $V_{\rm{LSR}}$ = $-$18$\pm$5 km s$^{-1}$ and $-$64$\pm$1 km s$^{-1}$ have been assigned for the source. Only $^{12}$CO (J=1$-$0) results are available for the source. Optically thick $^{12}$CO spectrum can often result in complex structures with multiple velocity components \citep{2007A&A...474..891U}.     
}\hss}}
\end{tabular} 
\end{center} 
\end{table*} 

\begin{table*}[htbp] 
\caption{Comparison between parallax-based and kinematic distances. \hspace{10em}}  
\begin{center}
\label{table:7} 
\small 
\begin{tabular}{lcccccc} 
\hline 
\hline 
 Target			&Parallax&Distance &\multicolumn{3}{c}{Kinematic distance\footnotemark[*]	}& Ref.\\
\cline{4-6}
	&$\pi$&$d_{\pi}$&$d_{V_{\rm{LSR}}}$&$d_{\mu_{l}}$	&$d_{3D}$						\\

	&(mas)&(kpc)&(kpc)&(kpc)	&(kpc)						\\
  \hline
G050.28-00.39					&0.140$\pm$0.018			&7.1$^{+1.1}_{-0.8}$	&9.3$^{+0.8}_{-0.9}$	&8.9$^{+3.0}_{-1.3}$	 &9.3$\pm$0.9	&1 		    		\\
G053.14+00.07					&0.726$\pm$0.038			&1.4$\pm$0.1&1.6$^{+0.6}_{-1.0}$	&$-$	 &$-$	&1 			\\
G060.57-00.18					&0.121$\pm$0.015			&8.3$^{+1.2}_{-0.9}$&9.5$\pm$1.0	&$-$	 &$-$	&2 		    	\\
G070.18+01.74					&0.136$\pm$0.014			&7.4$^{+0.8}_{-0.7}$&7.6$^{+1.1}_{-1.0}$	&6.6$^{+2.0}_{-2.5}$	 &7.4$\pm$0.9	 &3		    	\\

G070.33+01.59					&0.074$\pm$0.037			& \ \ \ $-$&7.5$^{+1.1}_{-1.2}$	&8.4$^{+1.9}_{-2.1}$	 &7.7$\pm$1.0&1	 		    		\\

G079.08+01.33\footnotemark[\dag]					&0.118$\pm$0.035			&8.5$^{+3.6}_{-1.9}$&9.1$^{+1.2}_{-1.1}$ 	&11.1$^{+1.6}_{-1.4}$	 &9.9$\pm$0.9 &1			    		\\

\hline 
\multicolumn{6}{@{}l@{}}{\hbox to 0pt{\parbox{120mm}{\normalsize
\par\noindent
\\
Column 1: Maser source; Columns 2-3: parallax and corresponding distance; Columns 4-5: Kinematic distances through LSR velocity and proper motion in the direction of Galactic longitude ($\mu_{l}$), respectively, where we assume the circular motion with a Universal rotation curve (i.e., A5 model of \citealp{2019ApJ...885..131R}); Column 6: Weighted mean of the both kinematic distances (Columns 4-5); Column 7: reference for the parallax result. \\
$\bf{References}$: (1) This paper; (2) \citet{2019ApJ...885..131R}; (3) \citet{2019AJ....157..200Z}.  \\
\footnotemark[*] We allowed deviation from circular motion for $>$13 km s$^{-1}$ in $V_{\rm{LSR}}$ and $\mu_{l}$ (at a source distance) (see red and black dotted lines in Fig. \ref{fig:2}).\\
\footnotemark[\dag] We apply $V_{\rm{LSR}}$ = $-$64$\pm$1 km s$^{-1}$ rather than $V_{\rm{LSR}}$ = $-$18$\pm$5 km s$^{-1}$ for G079.08+01.33 (see the text for details). 
}\hss}}
\end{tabular} 
\end{center} 
\end{table*}

\begin{figure*}[tbhp] 
 \begin{center} 
     \includegraphics[scale=1.0]{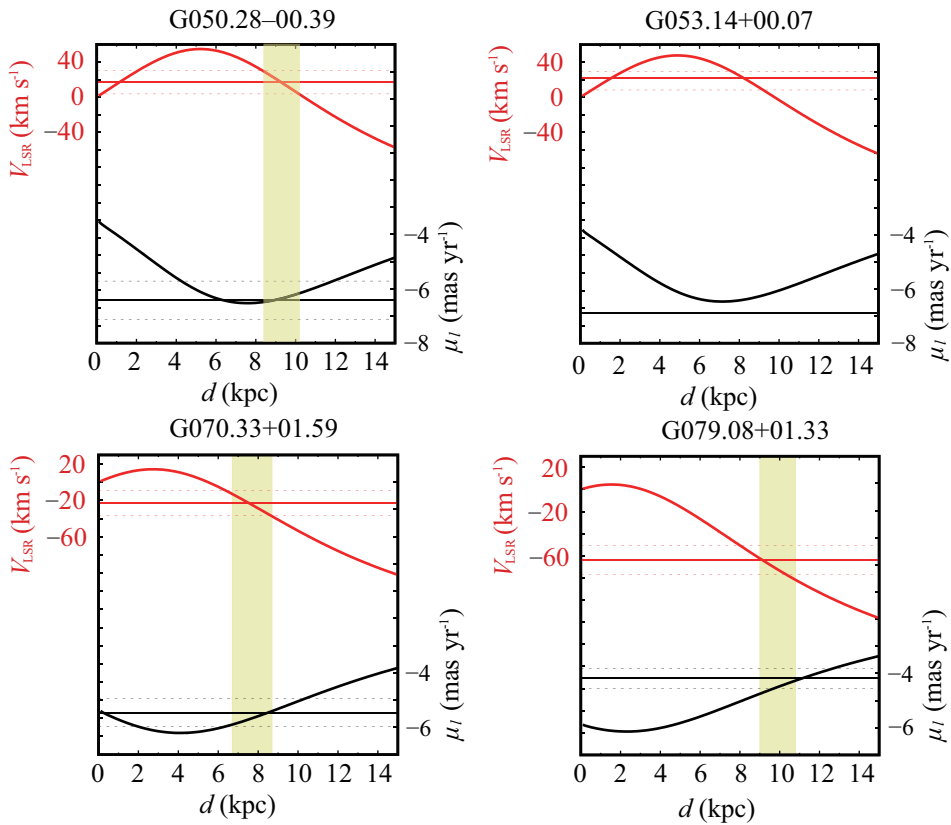} 
\end{center} 
\caption{Kinematic distance models through LSR velocity $V_{{\rm LSR}}$ (red curve; Y1 axis) and proper motion in the direction of Galactic longitude $\mu_{l}$ (black curve; Y2 axis). Horizontal axis is the heliocentric distance. Here, we assume the Galactic constants of ($R_{0}$ = 8.15 kpc, $\Theta_{0}$ = 236 km s$^{-1}$) and the circular motion with a universal rotation curve (A5 model of \citealp{2019ApJ...885..131R}). Horizontal red and black lines are observational results of $V_{{\rm LSR}}$ and $\mu_{l}$, respectively. Dotted red lines represent an error range where $\pm$13 km s$^{-1}$ is added in quadrature to a formal error of $V_{{\rm LSR}}$, while dotted black lines are the same as the red ones but for $\mu_{l}$ (see the text for details). Yellow shaded regions indicate a range for a weighted mean of both the kinematic distances (see Table \ref{table:7}). Regarding G079.08+1.33, $V_{{\rm LSR}}$ = $-$64$\pm$1 km s$^{-1}$ is adopted rather than $V_{{\rm LSR}}$ = $-$18$\pm$5 km s$^{-1}$ (see the text for details). }
\label{fig:2} 
\end{figure*}  

\section{Results}
\subsection{Parallax, proper motion, and 3D kinematic distance}
We measured trigonometric parallaxes and proper motions for four sources as shown in Fig. \ref{fig:1}, Table \ref{table:2}, and Appendix Table \ref{table:3}. We averaged the proper motion values in each source to assign a single proper motion (Appendix Table \ref{table:4}). We adopted these values for the central stars that excite the masers. For this uncertainty, we added $\pm$5 km s$^{-1}$ in quadrature to the fitted error estimates for each motion component. An uncertainty of 5 km s$^{-1}$ is consistent with the velocity dispersion of a typical molecular cloud (e.g., \citealp{1985ApJ...295..422C}). If only single proper motion was available, then an uncertainty of $\pm$10 km s$^{-1}$ was added (that is for G070.33+01.59). Thus, proper motion errors in Table \ref{table:2} are root-sum-square values of each formal error and 5 or 10 km s$^{-1}$. LSR velocities of the central stars were estimated by referring to molecular line observations (e.g., CO).

Based on LSR velocities ($V_{\rm{LSR}}$) and proper motions in the direction of Galactic longitude ($\mu_{l}$), we estimated 3D kinematic distances for individual sources as supplementary information (Table \ref{table:7} 
 and Fig. \ref{fig:2}). This is because we could not obtain a reliable parallax result for G070.33+01.59. We referred to \citet{2016PASJ...68...60Y} for the estimation of the 3D kinematic distances. We estimated kinematic distances through $V_{\rm{LSR}}$ and $\mu_{l}$ for each source by assuming circular motion with a Universal rotation curve (i.e., A5 model of \citealp{2019ApJ...885..131R}). The so-called near/far ambiguity was solved by comparing both the kinematic distances (i.e., $d_{V_{\rm{LSR}}}$ and $d_{\mu_{l}}$). Then, we determined the weighted mean of the both kinematic distances for each source (yellow regions in Fig. \ref{fig:2}) if available.

The kinematic distance estimation through either $V_{\rm{LSR}}$ or $\mu_{l}$, can be problematic in specific regions (e.g., $d_{V_{\rm{LSR}}}$ at Galactic longitudes 0$^{\circ}$ and 180${^{\circ}}$, and the solar circle; $d_{\mu_{l}}$ at the Sun$-$Galactic center line). Thus, reliable distance estimation can be achieved using both (e.g., see \citealp{2011PASJ...63..813S}). To conservatively estimate the uncertainty of the 3D kinematic distance, we allowed deviation from the circular motion for each motion component. The error range of $V_{\rm{LSR}}$ was determined as the root sum square of the formal error and 13 km s$^{-1}$ (red dotted lines in Fig. \ref{fig:2}). The uncertainty of 13 km s$^{-1}$ is consistent with the mean noncircular motion of the outer Perseus arm (\citealp{2019ApJ...876...30S}). The error range of $\mu_{l}$ was also determined with the same procedure (black dotted lines in Fig. \ref{fig:2}). To validate the obtained 3D kinematic distances, we expanded the determinations to G060.57-00.18 and G070.18+01.74 which have parallax results and are associated with the Perseus arm gap. In Table \ref{table:7}, differences between parallax-based and 3D kinematic distances, i.e. $|d_{\pi}$ $-$ $d_{3D}|$ / $\sqrt{\sigma_{d_{\pi}}^2+\sigma_{d_{3D}}^2}$ , range between 0$\sigma$ (for G070.18+01.74) and 1.5$\sigma$ (for G050.28-00.39).

\subsection{Individual sources}
Observational results of individual sources are shown here.

\subsubsection{G050.28-00.39 (IRAS 19230+1506)}

G050.28-00.39 is a site of current massive star formation \citep{2009ApJ...707..283J}, and is also defined as ultracompact (UC) H {\scriptsize II} region \citep{2003ApJ...587..714W}. By analyzing H$_{2}$CO absorption toward the UC H {\scriptsize II} region continuum emission, \citet{2003ApJ...587..714W} assigned a far kinematic distance of 10.0$\pm$0.8 kpc for the source.
Our parallax result corresponding to a distance of 7.1$^{+1.1}_{-0.8}$ (kpc), reveals that G050.28-00.39 is not within $\pm$3$\sigma$ of a global Perseus-arm model (see red circle in Figure \ref{fig:3}) although this source is classified as a Perseus-arm source in the $l-v$ plot of CO. The deviation from the global model indicates a spur of the Perseus arm, and this hypothesis should be validated by increasing astrometric results of the Perseus arm at Galactic longitude $l$$\sim$50$^{\circ}$ in the future.

\subsubsection{G053.14+00.07 (IRAS 19270+1750)}

While 1.1-mm continuum emission was detected in G053.14+00.07, continuum emissions at cm wavelengths have not been detected (\citealp{2012ApJ...755L..30A}). This is interpreted as a lack of H {\scriptsize II} region. Our parallax result corresponding to a distance of 1.4$\pm$0.1 (kpc), locates G053.14+0.07 in the interarm region between the Local and Sagittarius arms (see Fig. \ref{fig:3}). This prefers a near kinematic distance of the source (e.g., $d_{\rm{near}}$ =1.7 kpc adopted by \citealp{2018ApJ...863...74K}).

\subsubsection{G070.33+01.59 (K3-50)}
\begin{figure}[htbp] 
 \begin{center} 
     \includegraphics[scale=0.95]{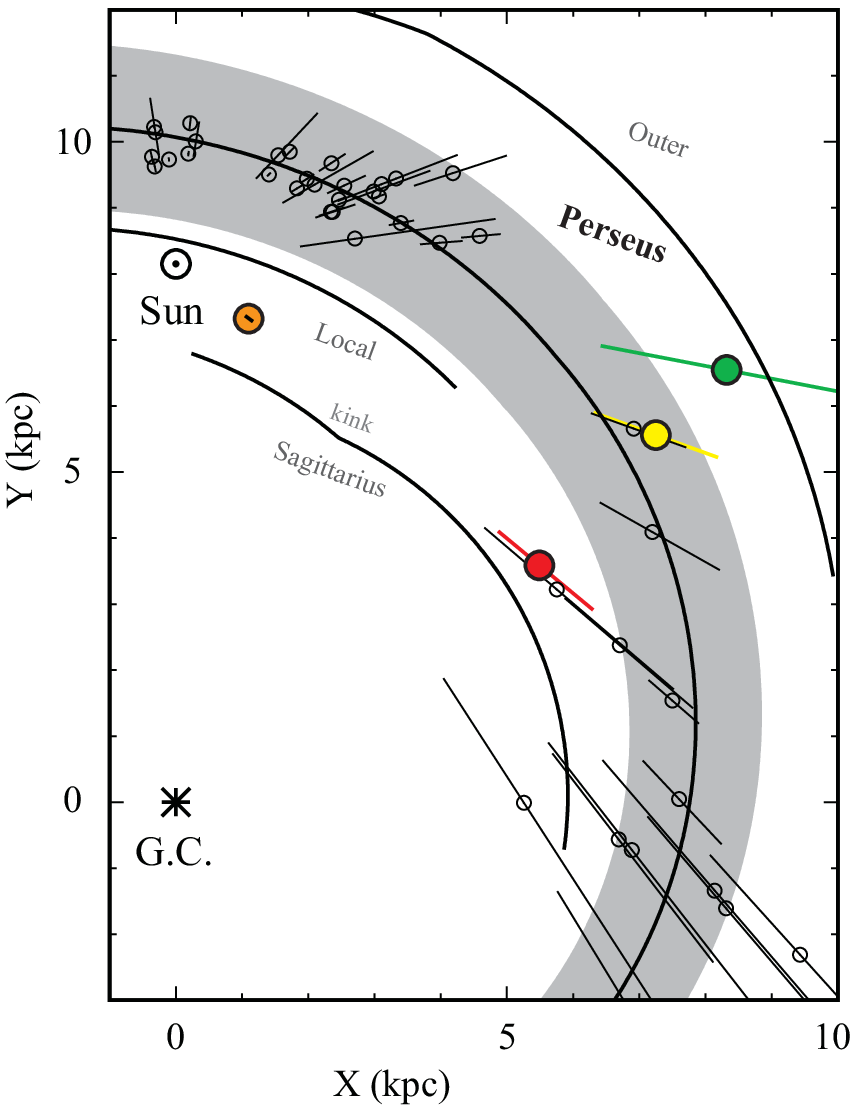} 
\end{center} 
\caption{ Face-on view of the Milky Way on which new astrometric results shown by colored circles with error bars, are superimposed (red = G050.28$-$00.39; orange = G053.14+0.07; yellow = G070.33+01.59; green = G079.08+01.33). Note that the 3D kinematic distance is applied for G070.33+01.59 (see Table \ref{table:7}). Previous VLBI astrometric results for the Perseus arm are plotted as open circles (\citealp{2019ApJ...885..131R}; \citealp{2020PASJ...72...50V}). The origin is the Galactic center, and the Sun is at ($X$, $Y$) = (0, 8.15) kpc, assumed. Black curves display log-periodic spiral-arm models of \citet{2019ApJ...885..131R} for the Outer, Perseus, Local and Sagittarius arms, respectively, from top to bottom. The gray shaded region indicates $\pm$3$\sigma$ of the arm width increasing with Galactocentric distance (see Fig.4 of \citealp{2019ApJ...885..131R}).}
\label{fig:3} 
\end{figure}

Several H {\scriptsize II} regions were identified by continuum observations at cm wavelengths (\citealp{2010ApJ...714.1015S}).
Although we succeeded in measuring the proper motion of G070.33+01.59, we could not obtain a reliable parallax (Table \ref{table:2}). By using the proper motion and LSR velocity, we estimated the 3D kinematic distance of the source to be 7.7$\pm$1.0 (kpc). The result locates the source in the Perseus arm gap (Fig. \ref{fig:3}). \citet{2019AJ....157..200Z} measured parallax and proper motion for G070.18+01.74 associated with the Perseus arm gap. The parallax-based distance of G070.18+01.74 ($d$ = 7.4$^{+0.8}_{-0.7}$ kpc) is consistent with a 3D kinematic distance of G070.33+01.59. In addition, both sources show the same LSR velocity and similar proper motions (i.e., the difference of the proper motions is within 1.5$\sigma$). Thus, G070.33+01.59 is likely associated with the Perseus arm gap.

\begin{table*}[htbp] 
\caption{VLBI astrometric results for the Perseus arm\footnotemark[$*$].} 
\begin{center} 
\label{table:5} 
\small 
\begin{tabular}{crrrrrrrc} 
\hline 
\hline 
 Target			&$V_{R}$ \ \ \ \ &$V_{\phi}$  \ \ \ \ &$V_{z}$   \ \ \ \ &$V_{\rm{LSR}}$   \ \ \  &$d$   \ \ \  &$R$   \ \ \ &$z$   \ \ \  &Ref. \ \ \ \ 			 \\

	&(km s$^{-1}$)&(km s$^{-1}$)&(km s$^{-1}$)	&(km s$^{-1}$)	&(kpc)&(kpc)&(pc)				\\
  \hline
G031.24-00.11 	&17$\pm$42		&231$\pm$60	&5$\pm$12&24$\pm$10 &13.2$^{+3.0}_{-2.0}$&7.5$^{+2.6}_{-1.6}$&$-$26$\pm$5&1\\
G032.79+00.19 	&$-$14$\pm$79			&172$\pm$52		&1$\pm$11&16$\pm$10&9.7$^{+4.2}_{-2.2}$&5.3$^{+3.1}_{-0.8}$&32$\pm$10&1\\
G037.50+00.53	&6$\pm$42			&200$\pm$38		&4$\pm$10&11$\pm$3&11.0$^{+2.3}_{-1.6}$&6.7$^{+1.8}_{-1.0}$&101$\pm$18&2\\
G037.82+00.41	&10$\pm$54			&212$\pm$51		&2$\pm$8&18$\pm$1&11.2$^{+3.3}_{-2.1}$&6.9$^{+2.6}_{-1.2}$&80$\pm$18&2\\
G040.42+00.70 	&2$\pm$40			&254$\pm$50		&14$\pm$6&10$\pm$5&12.8$^{+2.6}_{-1.8}$&8.5$^{+2.1}_{-1.3}$&156$\pm$26&1\\
G040.62-00.13 	&19$\pm$63			&261$\pm$75	&$-$3$\pm$9&31$\pm$3&12.5$^{+4.4}_{-2.6}$&8.2$^{+3.8}_{-1.8}$&$-$29$\pm$7&1\\
G042.03+00.19	 &8$\pm$39			&289$\pm$57		&$-$24$\pm$8&12$\pm$5&14.1$^{+2.9}_{-2.0}$&9.7$^{+2.5}_{-1.6}$&46$\pm$8&1\\
G043.16+00.01 	&5$\pm$20			&226$\pm$19	&11$\pm$11&10$\pm$5&11.1$^{+0.9}_{-0.8}$&7.6$^{+0.7}_{-0.5}$&1.9$\pm$0.2&1\\
G048.60+00.02	&9$\pm$13			&233$\pm$10	&6$\pm$6&18$\pm$5&10.0$\pm$0.5&7.7$\pm$0.3&3.5$\pm$0.2&1	\\
G049.26+00.31	&$-$15$\pm$34			&211$\pm$20	&$-$7$\pm$7&0$\pm$5&8.8$^{+1.5}_{-1.1}$&7.1$^{+0.8}_{-0.5}$&47$\pm$7&1\\
{\bf G049.41+00.32}&{\bf13$\pm$47}		&{\bf 160$\pm$17}	&{\bf31$\pm$14}&{\bf$-$21$\pm$1}&{\bf7.6$^{+2.3}_{-1.4}$}&{\bf 6.6$^{+1.1}_{-0.4}$}&{\bf42$\pm$10}&{\bf 1, a} \\
{\bf G050.28-00.39}&{\bf $-$7$\pm$}30		&{\bf 206$\pm$}10	&{\bf17$\pm$}18 	&{\bf17$\pm$3}&{\bf7.1$^{+1.1}_{-0.8}$}&{\bf6.6$^{+0.4}_{-0.2}$}&{\bf$-$49$\pm$6}&{\bf3, b}	    		\\
G060.57-00.18	&$-$18$\pm$28			&250$\pm$17&7$\pm$6		&4$\pm$5&8.3$^{+1.2}_{-0.9}$&8.3$^{+0.7}_{-0.4}$&$-$26$\pm$3&1 		    		\\
G070.18+01.74	&$-$14$\pm$19			&235$\pm$12&$-$4$\pm$6		&$-$23$\pm$5&7.4$^{+0.8}_{-0.7}$&8.9$^{+0.5}_{-0.3}$&223$\pm$23&4	    		\\

G094.60-01.79					&$-$5$\pm$7			&219$\pm$5&15$\pm$4			&$-$43$\pm$3&4.0$\pm$0.3	&9.4$^{+0.2}_{-0.1}$&$-$125$\pm$10&1	    		\\
G095.29-00.93					&6$\pm$3			&226$\pm$7&1$\pm$7			&$-$38$\pm$5&4.6$\pm$0.3	&9.7$\pm$0.2&$-$75$\pm$5&1	    		\\
G098.03+01.44					&$-$2$\pm$19			&187$\pm$10&27$\pm$16		&$-$61$\pm$5&2.7$^{+2.1}_{-0.8}$&8.9$^{+1.1}_{-0.3}$&68$\pm$30&1	    		\\
G100.37-03.57					&$-$13$\pm$11			&232$\pm$10&3$\pm$10			&$-$37$\pm$10&3.5$\pm$0.2	&9.4$\pm$0.1&$-$216$\pm$12&1	    		\\
G108.20+00.58					&11$\pm$12		&219$\pm$12&10$\pm$10	&$-$49$\pm$10&4.4$^{+0.9}_{-0.6}$&10.4$^{+0.6}_{-0.4}$&44$\pm$7&1	    		\\
G108.42+00.89					&$-$12$\pm$7			&210$\pm$6&0$\pm$6			&$-$51$\pm$5&2.5$^{+0.4}_{-0.3}$&9.2$^{+0.2}_{-0.1}$&38$\pm$5&1	    		\\
G108.47-02.81					&$-$20$\pm$7			&222$\pm$6&$-$9$\pm$7		&$-$54$\pm$5&3.2$\pm$0.1	&9.7$\pm$0.1&$-$159$\pm$5&1	    		\\
G108.59+00.49					&$-$46$\pm$7			&230$\pm$6&0$\pm$5			&$-$52$\pm$5&2.5$\pm$0.2	&9.2$\pm$0.1&21$\pm$2&1	    		\\

G110.19+02.47					&$-$5$\pm$10			&199$\pm$9&14$\pm$12		&$-$63$\pm$5&3.2$^{+0.9}_{-0.6}$&9.7$^{+0.6}_{-0.3}$&137$\pm$31&1	    		\\
G111.23-01.23					&$-$33$\pm$19			&238$\pm$17&$-$2$\pm$10		&$-$53$\pm$10&3.3$^{+1.2}_{-0.7}$&9.9$^{+0.8}_{-0.4}$&$-$72$\pm$19&1	    		\\
G111.25-00.76					&$-$1$\pm$5			&232$\pm$4&$-$3$\pm$6		&$-$40$\pm$3&3.6$\pm$0.2	&10.0$\pm$0.1&$-$48$\pm$3&1	    		\\
G111.54+00.77					&$-$18$\pm$5			&211$\pm$5&$-$10$\pm$5		&$-$57$\pm$5&2.6$\pm$0.1	&9.4$\pm$0.1&35$\pm$2&1	    		\\
G115.05-00.04					&$-$16$\pm$10			&241$\pm$9&$-$5$\pm$11		&$-$36$\pm$5&2.8$^{+0.4}_{-0.3}$&9.7$\pm$0.2&$-$2.0$\pm$0.2&1	    		\\

G122.01-07.08					&$-$30$\pm$5			&228$\pm$5&3$\pm$5			&$-$50$\pm$5&2.2$\pm$0.1	&9.4$\pm$0.1&$-$268$\pm$12&1	    		\\
G123.06-06.30					&$-$7$\pm$3			&239$\pm$3&$-$8$\pm$3		&$-$29$\pm$3&2.4$\pm$0.1	&9.6$\pm$0.1&$-$261$\pm$14&1	    		\\
G123.06-06.30					&$-$6$\pm$7			&242$\pm$7&$-$17$\pm$10		&$-$30$\pm$5&2.8$^{+0.3}_{-0.2}$&10.0$\pm$0.2&$-$310$\pm$26&1	    		\\
G133.94+01.06					&$-$21$\pm$3			&222$\pm$3&1$\pm$3			&$-$47$\pm$3&1.95$\pm$0.04			&9.61$\pm$0.04&36$\pm$1&1	    		\\
G134.62-02.19					&$-$3$\pm$5			&220$\pm$5&$-$6$\pm$6		&$-$39$\pm$5&2.4$\pm$0.1	&10.0$\pm$0.1&$-$93$\pm$4&1	    		\\

G136.84+01.16					&$-$6$\pm$5			&215$\pm$5&4$\pm$6			&$-$42$\pm$5&2.3$^{+0.9}_{-0.5}$&9.9$^{+0.7}_{-0.4}$&45$\pm$13&1	    		\\

G170.65-00.24					&$-$11$\pm$5			&221$\pm$8&$-$7$\pm$10		&$-$19$\pm$5&1.9$^{+0.3}_{-0.2}$&10.0$^{+0.3}_{-0.2}$&$-$8$\pm$1&1	    		\\
G173.48+02.44					&$-$9$\pm$5			&228$\pm$5&2$\pm$5			&$-$13$\pm$5&1.68$\pm$0.04			&9.82$\pm$0.04&71$\pm$2&1	    		\\
G174.20-00.07					&2$\pm$10		&244$\pm$7&9$\pm$6		&$-$2$\pm$10&2.1$\pm$0.1	&10.3$\pm$0.1&$-$2.6$\pm$0.1&1	    		\\

G183.72-03.66					&1$\pm$5			&236$\pm$10&4$\pm$10			&3$\pm$5&1.59$\pm$0.03			&9.73$\pm$0.03&$-$102$\pm$2&1	    		\\
G188.79+01.03					&$-$14$\pm$5			&213$\pm$8&$-$11$\pm$6		&$-$5$\pm$5&2.0$^{+0.5}_{-0.3}$&10.1$^{+0.5}_{-0.3}$&36$\pm$8&1	    		\\
G188.94+00.88					&0$\pm$5			&228$\pm$5&$-$4$\pm$6		&8$\pm$5&2.10$\pm$0.03			&10.23$\pm$0.03&32.3$\pm$0.4&1	    		\\
G192.16-03.81					&$-$3$\pm$5			&229$\pm$6&4$\pm$6			&5$\pm$5&1.5$\pm$0.1	&9.6$\pm$0.1&$-$101$\pm$6&1	    		\\
G192.60-00.04					&0$\pm$5			&240$\pm$5&5$\pm$5			&7$\pm$5&1.7$\pm$0.1	&9.8$\pm$0.1&$-$1.2$\pm$0.1&1	    		\\

G229.57-00.15					&$-$8$\pm$12			&218$\pm$14&$-$10$\pm$15		&53$\pm$10&4.6$^{+0.3}_{-0.2}$&11.7$\pm$0.2&12$\pm$1&1	    		\\

G236.81+01.98					&15$\pm$9		&226$\pm$9&$-$2$\pm$6	&53$\pm$10&3.1$^{+0.3}_{-0.2}$&10.2$\pm$0.2&105$\pm$8&1	    		\\

G240.31+00.07					&8$\pm$8			&231$\pm$9&6$\pm$12			&68$\pm$5&5.3$\pm$0.4	&11.8$\pm$0.3&6.5$\pm$0.5&1	    		\\

\hline 
\multicolumn{7}{@{}l@{}}{\hbox to 0pt{\parbox{150mm}{\normalsize
\par\noindent
\\
Column 1: Maser source; Columns 2-4: radial, azimuthal, and vertical velocities, respectively, in Galactocentric cylindrical coordinates. Azimuthal and vertical velocities are positive in the directions of the Galactic rotation and north Galactic pole, respectively; Column 5: LSR velocity; Column 6: heliocentric distance; Columns 7-8: Galactocentric radius and Galactic height, respectively; Column 9: reference. Number and alphabet indicate results of astrometry and a molecular line observation, respectively, if the both are listed. \\
$\bf{References}$: (1) \citet{2019ApJ...885..131R}; (2)\citet{2020PASJ...72...50V}; (3) This paper;  (4) \citet{2019AJ....157..200Z}; (a) \citet{2016ApJ...822...59S}; (b) \citet{2013ApJS..209....2S}. \\
\footnotemark[$*$] G049.41+00.32 and G050.28-00.39 are emphasized by bold font.
}\hss}}
\end{tabular} 
\end{center} 
\end{table*}

\begin{figure*}[tbhp] 
 \begin{center} 
     \includegraphics[scale=1.0]{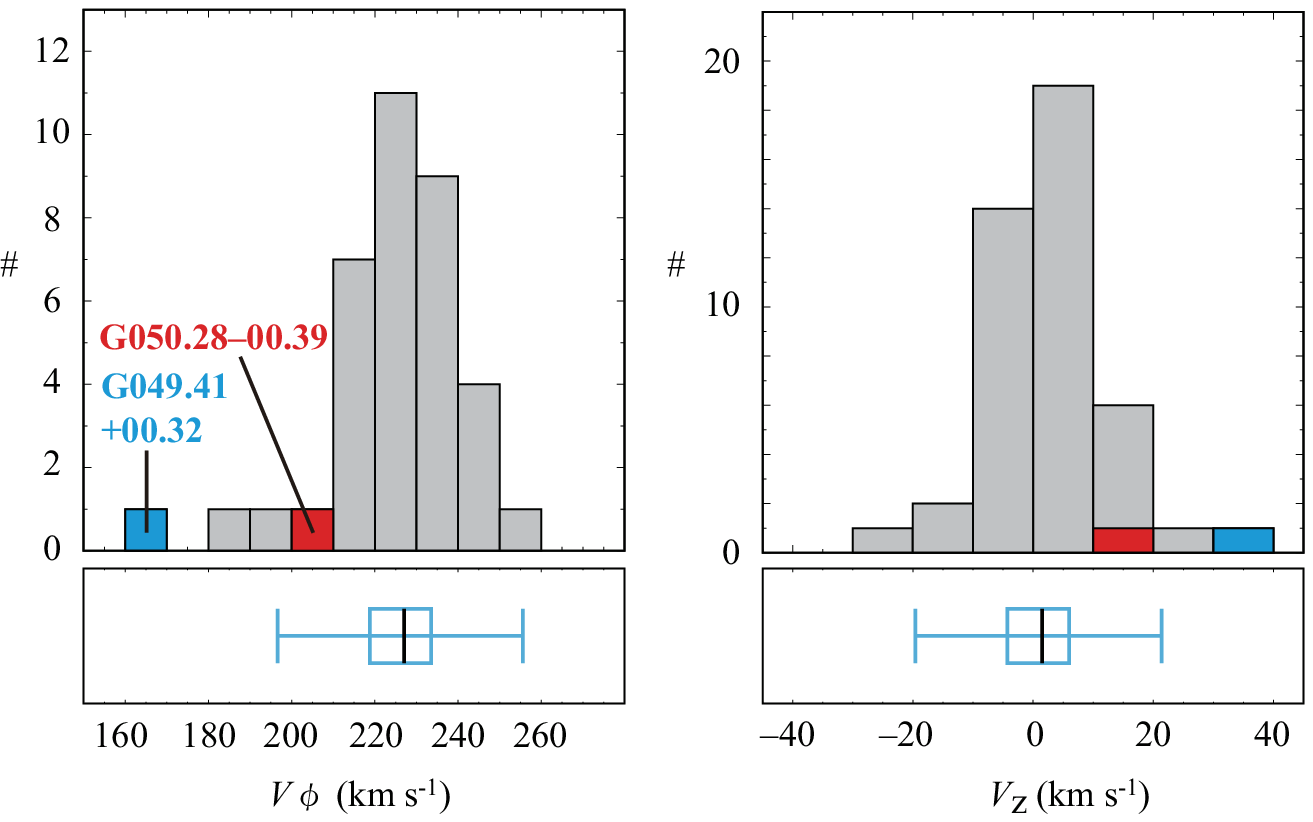} 
\end{center} 
\caption{\textbf{(Top row)} Histograms of $V_{\phi}$ ($Left$) and $V_{z}$ ($Right$) for the Perseus arm (referring to Table \ref{table:5}). Only sources whose motion uncertainties are less than 20 km s$^{-1}$, are plotted. Cyan and red emphasize G049.41+00.32 and G050.28-00.39, respectively. \textbf{(Bottom row)} Box plots of $V_{\phi}$ ($Left$) and $V_{z}$ ($Right$) are shown for the Perseus arm. The length of each whisker is 1.5$\times$IQR (interquartile range). The black vertical line in each plot shows the median value. }
\label{fig:8} 
\end{figure*}

\subsubsection{G079.08+01.33 (IRAS 20255+4032)}
This source was identified as compact H {\scriptsize II} region (\citealp{2011ApJ...727..114R}). Although a kinematic distance of 5 kpc was assigned for the source (e.g., \citealp{2002ApJS..141..157Y}), our result (i.e., 8.5$^{+3.6}_{-1.9}$ kpc) locates the source in the Outer arm rather than the Perseus arm (Fig. \ref{fig:3}). This discrepancy originates from the fact that different LSR velocities of  $V_{\rm{LSR}}$ = $-$18$\pm$5 km s$^{-1}$ and $V_{\rm{LSR}}$ = $-$64$\pm$1 km s$^{-1}$ have been applied for the source (e.g., \citealp{2002ApJS..141..157Y}; \citealp{2017ApJ...834...57M}). Only $^{12}$CO (J=1$-$0) results are available for the estimate of the source's LSR velocity. Ideally, the estimate with $^{12}$CO would be identical to that with dense gas tracers (e.g., $^{13}$CO, CS, NH$_{3}$, and HCO$^{+}$). However, $^{12}$CO often has multiple components along the line of sight, and makes the velocity estimate challenging \citep{2014ApJS..212....1A}. 

We apply $V_{\rm{LSR}}$ = $-$64$\pm$1 km s$^{-1}$ rather than $V_{\rm{LSR}}$ = $-$18$\pm$5 km s$^{-1}$ for G079.08+01.33 as explained below.
In Appendix Table \ref{table:4}, mean LSR velocities determined by maser observations of G050.28-00.39, G053.14+00.07, G070.33+01.59 and G079.08+01.33 are consistent with those determined by molecular line observations within $\sim$5 km s$^{-1}$ if we assume $V_{\rm{LSR}}$ = $-$64$\pm$1 km s$^{-1}$ for G079.08+01.33. Also, nearby Outer-arm sources G073.65+00.19 and G075.29+01.32 (listed in \citealp{2019ApJ...885..131R}) show LSR velocities of $-$76$\pm$10 km s$^{-1}$ and $-$58$\pm$5 km s$^{-1}$, respectively. Thus, $V_{\rm{LSR}}$ = $-$64$\pm$1 km s$^{-1}$ is preferable to $V_{\rm{LSR}}$ = $-$18$\pm$5 km s$^{-1}$ for G079.08+01.33.

\begin{figure*}[thpb] 
 \begin{center} 
     \includegraphics[scale=1]{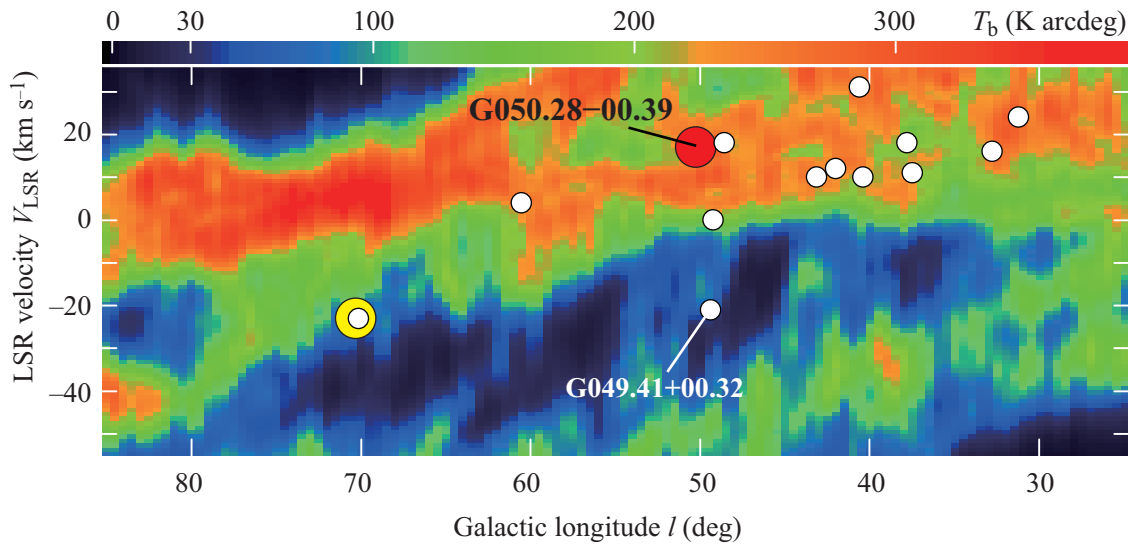} 
\end{center} 
\caption{ Galactic longitude vs. LSR velocity of H {\scriptsize I}, obtained by integrating the LAB data \citep{2005A&A...440..775K} over $\pm$0.5 degrees of Galactic latitude. The integration range covers G049.41+00.32 and G050.28$-$00.39. The color shows the integrated brightness temperature of the H {\scriptsize I} emission. Circles are Perseus-arm sources among which new astrometric results are shown by colored circles (i.e., red = G050.28-00.39; yellow = G070.33+01.59). We can see H {\scriptsize I} holes with brightness temperatures of less than 30 K arcdeg. }
\label{fig:6} 
\end{figure*}

\section{Discussion}
By combining our results with previous VLBI astrometric results (Table \ref{table:5}), we discuss the noncircular motion of the Perseus arm. We found that the noncircular motion is maximized near an end point of the Perseus arm gap as mentioned below. Three possible origins of the noncircular motion will be discussed. We argue that ``cloud collision with the Galactic disk'' is likely to explain VLBI and H {\scriptsize I} analysis results as shown below although further validations are necessary.

\subsection{Noncircular motion of the Perseus arm gap}

The noncircular motion of the Perseus arm is discussed based on sources whose motion uncertainties are less than 20 km s$^{-1}$ (Table \ref{table:5} and Fig. \ref{fig:8}). Here, the uncertainty of motion components ($V_{R}$, $V_{\phi}$, $V_{z}$) in Galactocentric cylindrical coordinates is estimated by considering errors of parallax, proper motion and LSR velocity, as summarized in the Appendix of  \citet{2020PASJ...72...53S}. 

Azimuthal velocities ($V_{\phi}$) of G049.41+00.32 and G050.28$-$00.39, 160$\pm$17 km s$^{-1}$ and 206$\pm$10 km s$^{-1}$, are smaller than the median value of the Perseus arm ($V_{\phi}$ = 227 km s$^{-1}$). If we subtract the universal rotation curve of \citet{2019ApJ...885..131R} from individual sources, residual azimuthal velocities of G049.41+00.32 and G050.28$-$00.39, $-$77$\pm$17 km s$^{-1}$ and $-$31$\pm$10 km s$^{-1}$, are still smaller than the median value of the Perseus arm ($-$6 km s$^{-1}$). 

\citet{2019AJ....157..200Z} listed two values of LSR velocity for G049.41+00.32 ($V_{\rm{LSR}}$ = $-$12$\pm$5 km s$^{-1}$ and $-$21$\pm$1 km s$^{-1}$), and we applied the latter value for the above discussion. Because the former and latter values come from CH$_{3}$OH maser and HCO$^{+}$ observations, respectively. The box and whisker plot for $V_{\phi}$ (Fig. \ref{fig:8} ($Bottom \ left$)) shows that G049.41+00.32 is clearly an outlier. 

Vertical velocities ($V_{z}$) of G049.41+00.32 and G050.28$-$00.39, 31$\pm$14 km s$^{-1}$ and 17$\pm$18 km s$^{-1}$ (Fig. \ref{fig:8} ($Right$)), may indicate systematic vertical motion perpendicular to the disk at an end point of the Perseus arm gap. Note that the vertical velocity is positive toward the north Galactic pole. Both sources are located near ($l$, $R$) = ($\sim$50$^{\circ}$, 6.6 kpc). The maximum and systematic noncircular motion is seen around an end point of the Perseus arm gap.   

Note that we cannot discuss radial velocities of G049.41+00.32 and G050.28$-$00.39 due to large uncertainties ($V_{R}$ = 13$\pm$47 km s$^{-1}$ and $-$7 $\pm$30 km s$^{-1}$).


\subsection{Possible origins of the noncircular motion}

Here, possible origins of the systematic noncircular motion of the Perseus arm gap (i.e., G049.41+00.32 and G050.28$-$00.39) are discussed.

\subsubsection{Spiral arm and/or Galactic bar}

The spiral arm and the Galactic bar are unlikely to be a physical origin of the noncircular motion. The noncircular motion of G049.41+00.32 ($>$ 70 km s$^{-1}$) is much greater than expected by the gravitational potential of the spiral arm (e.g., see Fig. 3 of \citealp{2019ApJ...876...30S}). Note that the determination of the noncircular motion of G049.41+00.32 is less affected by the internal motion of the maser source associated with G049.41+00.32. This is because \citet{2019AJ....157..200Z} estimated the systemic proper motion of the source based on observations of 6.7 GHz CH$_{3}$OH masers having internal motions of about 5 km s$^{-1}$ (e.g., \citealp{2010ApJ...716.1356M}). G049.41+00.32 and G050.28-00.39 are both located at a Galactocentric radius of 6.6 kpc, and this area might be less affected by the Galactic bar. Indeed, we examined 40 VLBI astrometric results (\citealp{2019ApJ...885..131R}; \citealp{2019A&A...632A.123I}) for the Scutum-Centaurus arm which is affected by the Galactic bar. The 40 sources are located at Galactocentric radii between 3.6 and 5.8 kpc, and the slowest azimuthal velocity is confirmed for G023.25-00.24 ($V_{\phi}$ = 182$\pm$17 km s$^{-1}$) at $R$ = 3.6 kpc.
The azimuthal velocity of G023.25-00.24 is consistent with that of G049.41+00.32 ($V_{\phi}$ = 160$\pm$17 km s$^{-1}$) within errors. However, G023.25-00.24 does not show vertical velocity perpendicular to the disk ($V_{z}$ = $-$3$\pm$7 km s$^{-1}$), which is inconsistent with that of G049.41+00.32 ($V_{z}$ = 31$\pm$14 km s$^{-1}$).\\

\subsubsection{Multiple supernova explosions}
It is difficult to explain the noncircular motion of the Perseus arm gap by star formation enhancement triggering supernova explosions, although we cannot conclusively reject the possibility. Star formation activity of the Perseus arm gap is weaker than the arm at $l >$ 90$^{\circ}$. No H {\scriptsize II} region is identified around G049.41+00.32. If G049.41+00.32 and G050.28$-$00.39 were affected by the same supernova explosions, we would find a hole in the Galactic longitude vs. LSR velocity of H {\scriptsize I} data as is seen for a Galactic superbubble (e.g., see Fig. 3 of \citealp{2008PASJ...60..975S}). To validate this hypothesis, we integrated the Leiden Argentina Bonn (LAB) H {\scriptsize I} survey data \citep{2005A&A...440..775K} over $\pm$0.5$^{\circ}$ of Galactic latitude. The integration range covers G049.41+00.32 and G050.28$-$00.39.

Fig. \ref{fig:6} shows that there are elliptical or rectangular H {\scriptsize I} holes with integrated brightness temperatures of less than 30 (K arcdeg). One of them is centered near ($l$, $V_{\rm{LSR}}$) = (47$^{\circ}$, $-$15 km s$^{-1}$) and G049.41+00.32 is associated with the rim of the hole. This result is consistent with typical characteristics of Galactic superbubbles. However, G050.28$-$00.39 is not associated with the rim of the H {\scriptsize I} hole. If the H {\scriptsize I} hole were created by multiple supernova explosions, we can estimate the expansion velocity of the H {\scriptsize I} hole along the line of sight. The range of LSR velocity of the H {\scriptsize I} hole is about [$-$25 km s$^{-1}$, $-$5 km s$^{-1}$], and thus the expansion velocity is 10 km s$^{-1}$. The estimated expansion velocity is much smaller than the noncircular motion of G049.41+00.32 (i.e., $>$ 70 km s$^{-1}$; see Fig. \ref{fig:8}). To conclusively accept or reject the possibility, star formation history of the Perseus arm gap should be studied in the future as was conducted for the stellar disk of the solar neighborhood in \citet{2019NatAs...3..932G}. \\


\begin{figure*}[htbp] 
 \begin{center} 
     \includegraphics[scale=1]{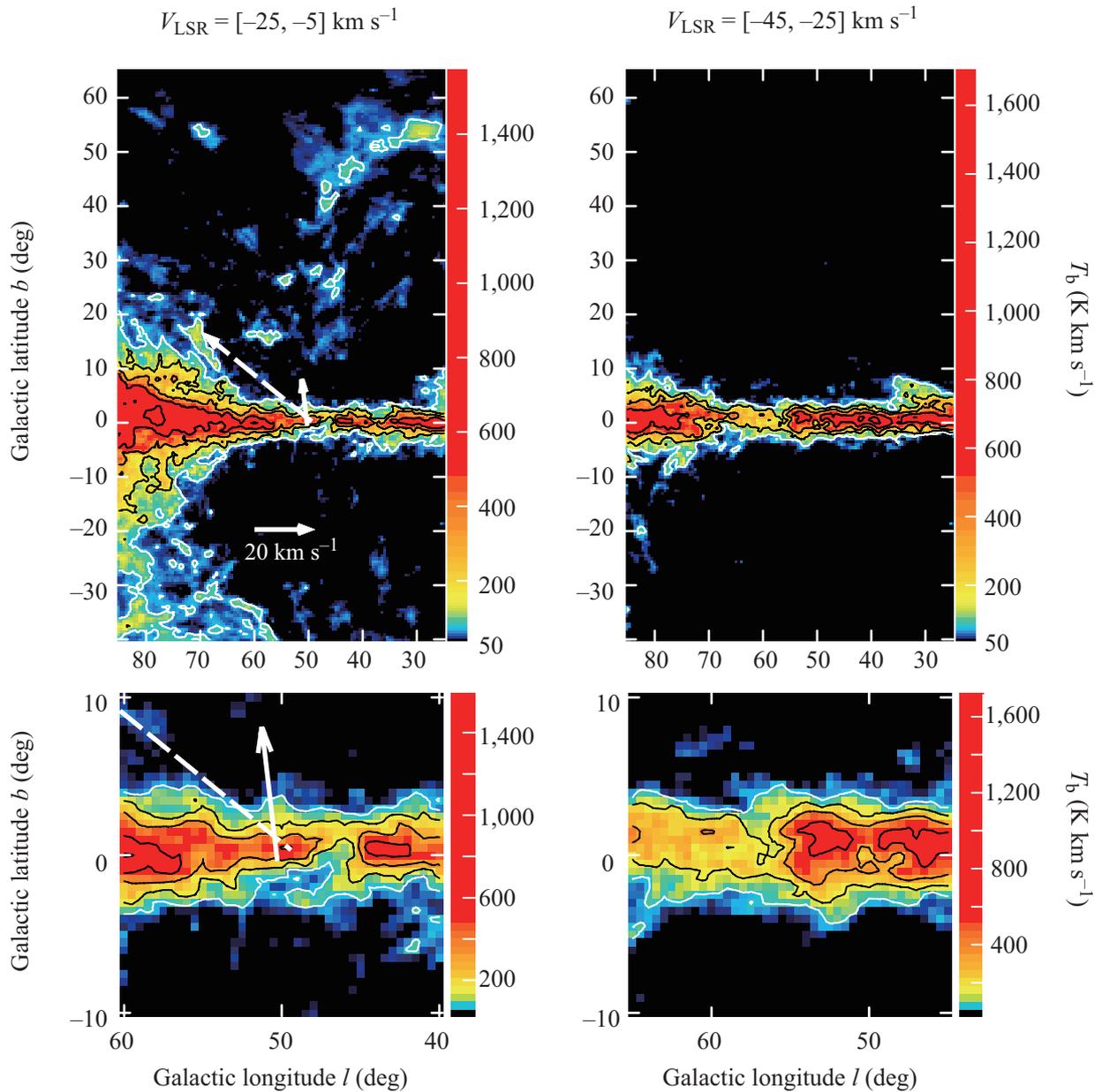} 
\end{center} 
\caption{ \textbf{(Top row)} Velocity-integrated H {\scriptsize I} maps are shown in Galactic coordinates. Color and contours represent the integrated brightness temperature in the unit of K km s$^{-1}$. The contour levels are 5$\%$, 10$\%$, 20$\%$, 40$\%$ and 80$\%$ of a peak brightness temperature ($T_{b}$ = 1563 and 1693 K km s$^{-1}$ for $Left$ and $Right$ panels, respectively). The velocity integration ranges are  $V_{\rm{LSR}}$ = [$-$25 km s$^{-1}$, $-$5 km s$^{-1}$] ($Left$) and [$-$45 km s$^{-1}$, $-$25 km s$^{-1}$] ($Right$). Motion vectors corrected for the universal rotation curve of \citet{2019ApJ...885..131R}, are plotted for G049.41+00.32 (dashed arrow) and G050.28-00.39 (solid arrow), respectively (i.e., noncircular motion vectors). The end points of individual arrows are consistent with the sources' positions. \textbf{(Bottom row)} Enlarged display for the top panels. }
\label{fig:12} 
\end{figure*}  

\subsubsection{Cloud collision with the Galactic disk}

A cloud collision with the Galactic disk could explain the noncircular motion of the Perseus arm gap. The large Galactocentric distance of G049.41+00.32 and G050.28-00.39 (i.e., $R$ = 6.6 kpc) as well as the systematic vertical motion perpendicular to the disk could be naturally explained by the cloud collision. To quantitatively confirm whether slow azimuthal velocities of the sources can be explained by the cloud collision, a simulation study should be conducted in the future.

The $l$-$v$ plot of H {\scriptsize I} (Fig. \ref{fig:6}) shows elliptical or rectangular holes where the LSR velocity ($V_{\rm{LSR}}$) of the holes is gradually decreased with an increasing Galactic longitude ($l$). Thus, we integrated the H {\scriptsize I} data over two velocity ranges, $V_{\rm{LSR}}$ = [$-$25, $-$5 ] km s$^{-1}$ and [$-$45, $-$25] km s$^{-1}$. The resultant maps are shown in Galactic coordinates (Fig. \ref{fig:12}).
 
We can see extended H {\scriptsize I} emission on one side of the Galactic plane (i.e., above the plane) at the former velocity range $V_{\rm{LSR}}$ = [$-$25, $-$5] km s$^{-1}$. G049.41+00.32 is included in the velocity range. G049.41+00.32 and G050.28-00.39 are moving toward the extended H {\scriptsize I} emission, indicating that the sources are physically related to the H {\scriptsize I} emission although the distance of the H {\scriptsize I} emission is unknown. The Galactic H {\scriptsize I} disk shows an arc shape near Galactic longitude $l$ = 46$^{\circ}$ in Fig. \ref{fig:12} ($Lower \ left$), which suggests that the disk was pushed from the lower side of the disk. To estimate the total mass of the extended H {\scriptsize I} gas, we integrated Fig. \ref{fig:12} ($Upper \ left$) over ranges of Galactic longitude $l$ = [25$^{\circ}$, 85$^{\circ}$] and latitude $b$ = [40$^{\circ}$, 60$^{\circ}$]. The integrated H {\scriptsize I} mass is scaled as
\begin{eqnarray}
\label{Eq:2}
   M_{\rm{H I}} \ [\rm{10}^{6} \ M_{\odot}] & = & 7.8 \times \left( \frac{d}{\rm{7.6 \ kpc}}  \right)^{2} \rm{above \ the \ disk}
\end{eqnarray}
 where $M_{\odot}$ is the solar mass, $d$ is heliocentric distance, and 7.6 kpc is the distance of G049.41+00.32. In Eq. (\ref{Eq:2}), we assume that (1) H {\scriptsize I} gas within the restricted range is located at the same distance and (2) H {\scriptsize I} emission is optically thin with a conversion factor $X$ of 1.82 $\times$10$^{18}$ [H cm$^{-2}$ K$^{-1}$ km$^{-1}$ s] (\citealp{2001ApJ...547..792D}). The mean column density is 7.0$\times$10$^{19}$ cm$^{-2}$ in Eq. (\ref{Eq:2}) while the maximum value detected at ($l$, $b$) = (29$^{\circ}_{.}$0, 53$^{\circ}_{.}$5), is 2.4$\times$10$^{20}$ cm$^{-2}$. 

Eq. (\ref{Eq:2}) indicates that a cloud comparable to the Smith cloud with a H {\scriptsize I} mass of $>$10$^{6}$ $M_{\odot}$, collided with the disk if our assumption of the cloud collision is correct. Indeed, the maximum column density of the Smith cloud, $\sim$5$\times$10$^{20}$ cm$^{-2}$ (\citealp{2008ApJ...679L..21L}), is comparable to that of the extended H {\scriptsize I} gas. \citet{2018MNRAS.473.5514T} proposed that the Smith cloud is confined by a DM subhalo with a mass of $>$10$^{8}$ $M_{\odot}$. Based on a cosmological simulation by \citet{2004MNRAS.355..819G}, we can see a ``Smith-sized halo'' every 0.1$-$1 Gyr inside the optical disk of the Galaxy. The discovery rate is inconsistent with our assumption of the cloud collision with the disk if the cloud is comparable to the Smith cloud. 

We cannot see a similar extended emission at the latter velocity range $V_{\rm{LSR}}$ = [$-$45, $-$25] km s$^{-1}$ (see Fig. \ref{fig:12} ($Upper \ right$)). Instead, H {\scriptsize I} emission in the disk is weakened at the Galactic longitude range $l$ $\sim$ [55$^{\circ}$, 65$^{\circ}$]. It might be explained by assuming that a H {\scriptsize I} hole created by the cloud collision was sheared by the differential Galactic rotation. \\

{\bf
We have discussed three possible origins, among which ``the cloud collision with the Galactic disk'' best matches the observational results. However, further investigations are necessary in the future. For example,  
 distance estimation of the extended H {\scriptsize I} emission via an absorption line observation would be valuable. We should also examine the star formation history of the Perseus arm gap based on stellar data as mentioned previously. }

\section{Summary}
The main results in this paper are summarized as follows.

\begin{enumerate}
   \item We reported VERA astrometric results (Figure \ref{fig:1} and Table \ref{table:2}) for G050.28-00.39, G053.14+00.07, G070.33+01.59, and G079.08+01.33. Since trigonometric parallax was not accurately measured for G070.33+01.59, we estimated the source's distance using LSR velocity and the measured proper motion (Fig. \ref{fig:2} and Table \ref{table:7}). G050.28-00.39 and G070.33+01.59 are likely associated with the Perseus arm gap (Figure \ref{fig:3}). G053.14+00.07 is a nearby star-forming region while G079.08+01.33 is associated with the Outer arm. Still, the number of VLBI astrometric results for the Perseus arm gap is smaller compared to the other part of the arm. VLBI astrometric results support the previous discussion that there is a lack of raw material for present and future star formation in the Perseus arm gap.  \\

   \item The noncircular motion of the Perseus arm is maximized near Galactic longitude $l\sim$50$^{\circ}$ where G049.41+00.32 and G050.28-00.39 lag relative to the Galactic rotation by 77$\pm$17 km s$^{-1}$ and 31$\pm$10 km s$^{-1}$, respectively (Fig. \ref{fig:8}; Table \ref{table:5}). The sources' location is near an end point of the Perseus arm gap. The noncircular motion of G049.41+00.32 cannot be explained by the gravitational potential of the spiral arm, and G049.41+00.32 and G050.28-00.39 are moving toward the north Galactic pole with $V_{z}$ = 31$\pm$14 km s$^{-1}$ and 17$\pm$17 km s$^{-1}$, respectively. Because the Galactocentric radii of G049.41+00.32 and G050.28-00.39 are $R$ = 6.6$^{+1.1}_{-0.4}$ kpc and 6.6$^{+0.4}_{-0.2}$ kpc, respectively, the sources might be less affected by the Galactic bar. \\

 \item To understand the origin of the noncircular motion of G049.41+00.32 and G050.28-00.39, we analyzed the LAB H {\scriptsize I} data (Figures \ref{fig:6}-\ref{fig:12}). We found elliptical and/or rectangular holes in the $l-v$ plot of H {\scriptsize I} (Fig. \ref{fig:6}). G049.41+00.32 is associated with the rim of a hole whereas G050.28-00.39 is not associated with the hole. We integrated the H {\scriptsize I} data over velocities covering the holes. We did not find any Galactic superbubbles in Galactic coordinates (Fig. \ref{fig:12}). Instead, extended H {\scriptsize I} emission was seen on one side of the Galactic plane at the velocity range $V_{\rm{LSR}}$ = [$-$25 km s$^{-1}$, $-$5 km s$^{-1}$], and the Galactic H {\scriptsize I} disk showed an arc feature at the same velocity range. G049.41+00.32 and G050.28-00.39 are moving toward the extended H {\scriptsize I} emission. These observational results could be explained by ``a cloud collision with the Galactic disk'' although further investigations are needed (e.g., distance estimation of the extended H {\scriptsize I} emission via an absorption line observation). \\

\end{enumerate}

\begin{ack}
We deeply acknowledge Dr. Takumi Nagayama as well as VERA project members for their supports on VERA observations and data reduction. We thank the anonymous referee for valuable comments which improved the manuscript. We would like to thank Editage (www.editage.com) and Prof. Juli Scherer for English language editing. Data analysis was in part carried out on common use data analysis computer system at the Astronomy Data Center, ADC, of NAOJ.


\end{ack}


\bibliographystyle{apj}
\bibliography{reference_pasj}



\appendix
\section{Supplemental materials}
\begin{figure*}[tbhp] 
 \begin{center} 
     \includegraphics[scale=1.4]{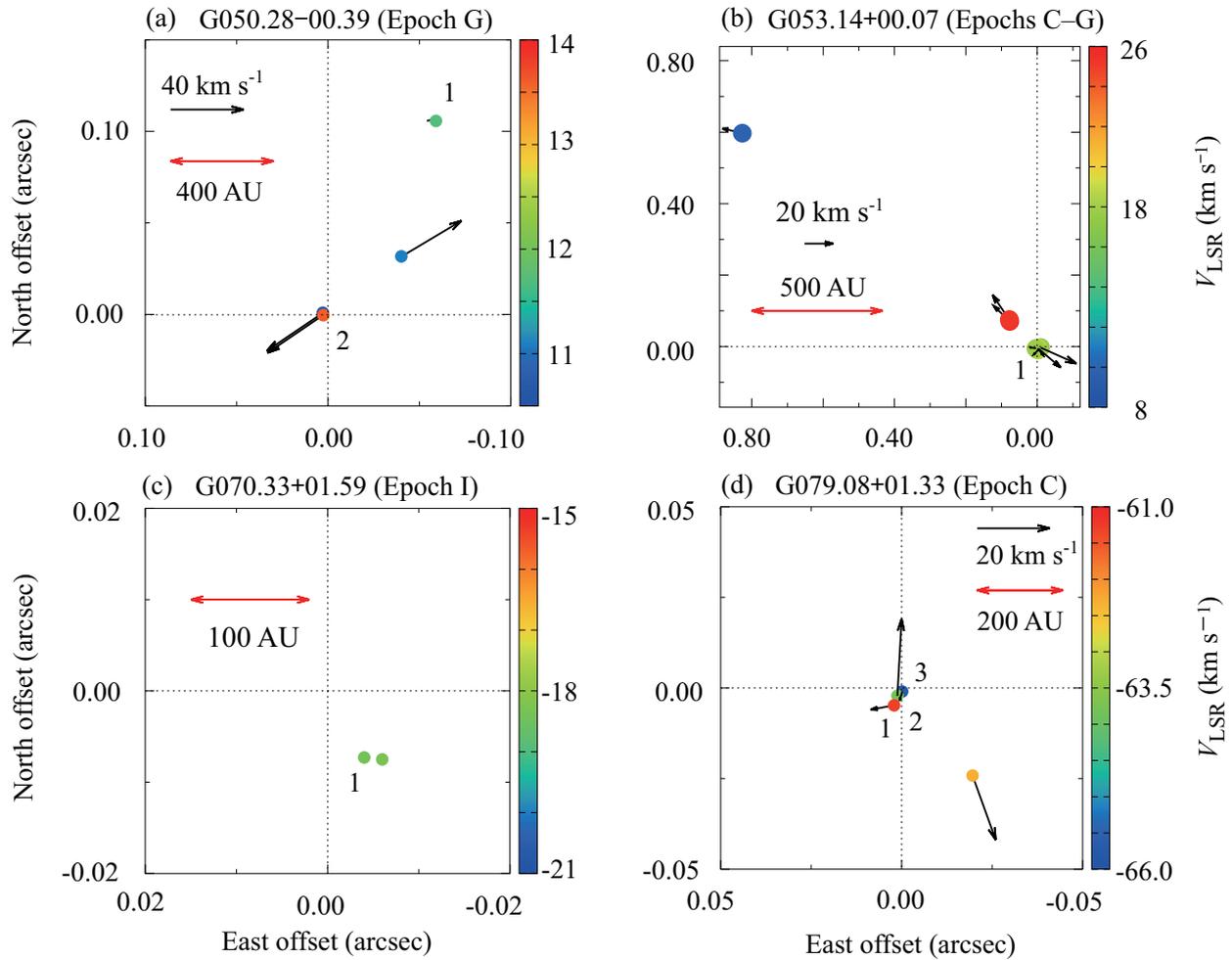} 
\end{center} 
    \caption{Distribution maps of maser features are shown with internal motion vectors for {\bf (a)} G050.28$-$00.39, {\bf(b)} G053.14+00.07,
{\bf(c)} G070.33$+$01.59, and {\bf(d)} G079.08+01.33. 
The origin of  each map (coordinates) is described in Table \ref{table1}. The parenthesis at the top of each map indicates reference epoch(s) (see Table \ref{table:4}). Maser features used for the parallax fit, are labeled with feature numbers defined in Table \ref{table:3}.
The horizontal red arrow in each map shows
a spatial scale converted at a source distance (see Tables \ref{table:2} and \ref{table:7}). A horizontal black arrow displays a scale of absolute velocity. Color bar indicates LSR velocity. }    
     \label{fig:5} 
\end{figure*}

Here, we show supplemental materials to further document
parallaxes (table \ref{table:3}), proper motions
(table \ref{table:4}), and detailed maser maps (figure \ref{fig:5}).



\begin{table*}
\caption{Trigonometric parallaxes for individual sources\footnotemark[$*$].}
\label{table:3}
\begin{center}
\begin{tabular}{cccccccc}
\hline
\hline
Source&Feature	&$V_{\rm{LSR}}$	&Detection	&Parallax	&	&\multicolumn{2}{c}{Errors} \\
 \cline{7-8} 
&		&km s$^{-1}$	&in Epoch\footnotemark[$\dag$]							&(mas)	&	&R.A.	&Decl. \\
&		&	&				&				&	&(mas)&(mas) \\  
\hline
G050.28-00.39 &1	&11.0$\sim$12.3		&$\times$BCDEFGHI$\times$		&0.141$\pm$0.036		&	&0.067		&0.140	\\
(J1924+1540)				  &2	&12.3$\sim$14.8		&$\times$$\times$$\times$DEFGHIJ		&0.163$\pm$0.020		&	&0.038		&0.047	\\
G050.28-00.39 &1	&11.0$\sim$12.3		&$\times$BCD$\times$FGHIJ		&0.093$\pm$0.027		&	&0.048		&0.058	\\
(J1930+1532)	&2	&12.3$\sim$14.8		&$\times$$\times$$\times$DEFGHIJ						&0.144$\pm$0.022		&	&0.040		&0.062\\
\hline
\multicolumn{3}{l}{Combined fit}							&0.140$\pm$0.012	&		&	&		\\
\multicolumn{3}{l}{Final result}							&0.140$\pm$0.018	&		&	&		\\ \\

G053.14+00.07 &1	&15.7$\sim$18.4		&AB$\times$DEFGHIJ		&0.733$\pm$0.063		&	&0.130		&0.128	\\
(J1927+1847)	&	&		&		&		&	&		&		\\ 
G053.14+00.07 &1	&15.7$\sim$18.4		&ABCDEFGHIJ		&0.722$\pm$0.050		&	&0.109		&0.138	\\
(J1928+1842)	&	&		&		&		&	&		&		\\ 
\hline
\multicolumn{3}{l}{Combined fit}							&0.726$\pm$0.038	&		&	&		\\
\multicolumn{3}{l}{Final result}							&0.726$\pm$0.038	&		&	&		\\ \\

G070.33+01.59 &1	&$-$18.8$\sim$$-$17.8		&A$\times$CDEFGHI$\times$		&0.096$\pm$0.047		&	&0.141		&0.089	\\
(J1957+3338)	&	&		&	&		&	&		&		\\ 
G070.33+01.59 &1	&$-$18.8$\sim$$-$18.0		&A$\times$CDEFGH$\times$$\times$		&0.039$\pm$0.062		&	&0.122		&0.175\\
(J2010+3322)	&	&		&	&		&	&		&		\\ 
\hline
\multicolumn{3}{l}{Combined fit}							&0.074$\pm$0.037	&		&	&		\\
\multicolumn{3}{l}{Final result}							&0.074$\pm$0.037	&		&	&		\\ \\

G079.08+01.33 &1	&$-$61.7$\sim$$-$60.8		&ABCDEFG$\times$$\times$		&0.081$\pm$0.048		&	&0.094		&0.104	\\
(J2033+4000)				  &2	&$-$64.6$\sim$$-$63.0		&ABCDEFGHI		&0.110$\pm$0.026		&	&0.057		&0.107	\\
				  &3	&$-$65.9$\sim$$-$65.5		&ABCD$\times$FGHI		&0.172$\pm$0.042		&	&0.098		&0.096	\\
\hline
\multicolumn{3}{l}{Combined fit}							&0.118$\pm$0.020	&		&	&		\\
\multicolumn{3}{l}{Final result}							&0.118$\pm$0.035	&		&	&		\\ \\

\hline

\hline

\multicolumn{4}{@{}l@{}}{\hbox to 0pt{\parbox{140mm}{\footnotesize
\par\noindent
\\
Column 1: 22 GHz H$_{2}$O maser source (as denoted by ``G") and background
QSOs (as denoted by ``J"); Column 2: feature ID; Column 3: range of $V_{\rm{LSR}}$; Column 4: detection flag ($\times$ = non detection); Column 5: parallax result; Columns 6-7: astrometric errors in R.A. and Decl., respectively.\\ 
\footnotemark[$*$]Positional errors were adjusted so that the reduced chi-square values become unity. Since the atmospheric delay residual is common among different maser features, the error of the final parallax is multiplied by the number of maser features. \\
\footnotemark[$\dag$] Please refer to Table \ref{table1} for the corresponding epoch date. 
}\hss}}
\end{tabular}
\end{center}
\end{table*}


\begin{table*}[tbp]
\caption{Systemic proper motions of star-forming regions.}
\label{table:4}
\begin{center}
\begin{tabular}{ccccccrrc}
\hline
\hline
Source&Feature	&	&\multicolumn{2}{c}{Proper Motion\footnotemark[$\ast$]}&&\multicolumn{2}{c}{Position offset\footnotemark[$\dag$]}&Ref.	\\
\cline{4-5} \cline{7-8}
&	&$V_{\rm{LSR}}$	&$\mu_{\alpha} \rm{cos}\delta$	&$\mu_{\delta}$	&&$\Delta \alpha$&$\Delta \delta$&epoch\\
&	&km s$^{-1}$	&(mas yr$^{-1}$)&(mas yr$^{-1}$)&&(mas)&(mas)	\\
\hline
G050.28-00.39&1 	&11.0$\sim$12.3 	&$-$3.23$\pm$0.04	&$-$5.50$\pm$0.05&&$-$59&106	&G	\\ 
&2	$\&$ 2b	&12.3 $\sim$ 14.8 $\&$ 10.2 $\sim$ 11.0	&$-$2.39$\pm$0.04	&$-$6.13$\pm$0.05&&3&1		&G\\
&3			&10.6 $\sim$ 11.4 	&$-$4.26$\pm$0.23	&$-$4.94$\pm$0.20	&&$-$40&32	&G\\
\hline
&Unweighted mean		&11.8				 	&$-$3.29$\pm$0.54\footnotemark[$\ddag$]	&$-$5.52$\pm$0.34\footnotemark[$\ddag$]		\\ \\

G053.14+00.07&1 	&15.7 $\sim$ 18.4	&$-$5.09$\pm$0.10	&$-$9.43$\pm$0.11&&$-$10&$-$3	&D	\\
&2			&18.6 $\sim$ 20.9 	&$-$3.64$\pm$0.35	&$-$9.82$\pm$0.22&&$-$4&$-$7	&D	\\

&3			&25.2 $\sim$ 26.2 	&0.45$\pm$0.85	&$-$4.95$\pm$1.08&&77&71	&C	\\

&4			&9.4 $\sim$ 10.4 	&0.86$\pm$0.02	&$-$6.49$\pm$0.05&&826&598	&D	\\
&5			&17.4 $\sim$ 18.2 	&$-$0.59$\pm$0.76	&$-$7.01$\pm$0.86&&3&$-$5&F		\\
&6			&18.4 $\sim$ 19.0 	&$-$0.84$\pm$0.46	&$-$7.73$\pm$0.52&&$-$2&$-$10		&F\\
&7			&18.4 $\sim$ 19.3 	&$-$1.84$\pm$0.33	&$-$8.13$\pm$0.01&&$-$1&$-$5	&E	\\
&8			&24.7 $\sim$ 25.8 	&0.53$\pm$0.70	&$-$3.64$\pm$0.81&&78&75	&G	\\

\hline
&Unweighted mean		&19.1		&$-$1.27$\pm$0.75\footnotemark[$\ddag$]	&$-$7.15$\pm$0.75\footnotemark[$\ddag$]		\\ \\

G070.33+01.59&1 $\&$ 1b 			&$-$19.0 $\sim$ $-$17.6	&$-$2.82$\pm$0.08	&$-$4.68$\pm$0.06	&&$-$5&$-$8	&I\\ \\

G079.08+01.33&1 			&$-$61.7 $\sim$ $-$60.8 	&$-$2.33$\pm$0.09	&$-$3.39$\pm$0.10&&2&$-$5	&C	\\ 
&2			&$-$64.6 $\sim$ $-$63.0 	&$-$2.52$\pm$0.04	&$-$2.83$\pm$0.07	&&1&$-$2&C	\\
&3			&$-$65.9 $\sim$ $-$65.5 	&$-$2.47$\pm$0.06	&$-$3.42$\pm$0.06	&&0&$-$1&C	\\
&4			&$-$62.5 $\sim$ $-$62.1 	&$-$2.65$\pm$0.23	&$-$3.80$\pm$0.46	&&$-$20&$-$24	&C\\
\hline
&Unweighted mean		&$-$63.3				 	&$-$2.49$\pm$0.07\footnotemark[$\ddag$]	&$-$3.36$\pm$0.20\footnotemark[$\ddag$]		\\ 
\hline

\multicolumn{4}{@{}l@{}}{\hbox to 0pt{\parbox{170mm}{\footnotesize
\par\noindent \\
Column 1: source name; Column 2: feature ID same as Table \ref{table:3}; Column 3: range of $V_{\rm{LSR}}$; Columns 4-5: proper motion values in R.A. and Decl., respectively; Columns 6-7: position offsets in R.A. and Decl., respectively, relative to individual source coordinates listed in Table \ref{table1}; Column 8: reference epoch for the position offset (see the corresponding epoch in Table \ref{table1}). \\
\footnotemark[$\ast$]
Proper motion components are determined by adapting a final distance (see Table \ref{table:2}) or 3d kinematic distance (that is for G070.33+01.59).  \\
\footnotemark[$\dag$]
A mean position of different maser spots which are close to each other and show similar LSR velocities, is displayed for each maser feature. This is because the maser spots are correlated. \\
\footnotemark[$\ddag$]
An error is the standard error. \\

}\hss}}

\end{tabular}
\end{center}
\end{table*}

\end{document}